\documentstyle[11pt,graphics]{article}
\def\reference{\parskip 0pt\par\noindent\hangindent 0.5 truecm}

\begin{document}
%
%
\title{The Hubble Constant from (CLASS) Gravitational Lenses}
%



\author{L.V.E. Koopmans$^{1}$ + the CLASS collaboration
} 

\date{}
\maketitle

{\center $^1$ Caltech, mailcode 130-33, Pasadena CA 91125, USA \\
leon@tapir.caltech.edu\\[3mm]}

\begin{abstract}
\noindent
One of the main objectives of the {\sl Cosmic Lens All-Sky Survey}
(CLASS) collaboration has been to find gravitational lens (GL) systems
at radio wavelengths that are suitable for the determination of
time delays between image pairs. The survey is now near completion and
at least 18 GL systems have been found. Here, I will discuss our
efforts to measure time delays from several of these systems with the
ultimate aim of constraining the Hubble Constant~(H$_0$). Thus far
three CLASS GL systems (i.e. B0218+357, B1600+434 and B1608+656) have
yielded measurements of time delays, from which values of
H$_0$$\approx$60--70~km\,s$^{-1}$\,Mpc$^{-1}$ have been
estimated. Although most GL systems give similar values of H$_0$,
statistical {\sl and} systematic uncertainties are still considerable.
To reduce these uncertainties, I will shortly mention two monitoring
programs that we are undertaking to (re)measure time delays in 15
CLASS GL systems and address several important issues for the
near future.
\end{abstract}


\bigskip

\section{The Hubble Constant from Gravitational Lensing}

Refsdal (1964) showed that to first order the Hubble Constant can be
measured from a multiple-image GL system, if the time delay between an
image pair and the mass distribution of the deflector is known. This
has prompted the monitoring of and search for new GL systems, after
the discovery of the first GL system Q0957+561 (Walsh et
al. 1979). Only recently has the time delay in Q0957+561 been measured
unambiguously (e.g. Kundic et al. 1997). Since then time delays from
seven other GL systems have been reported, of which five (including
Q0957+561) have 1--$\sigma$ time-delay errors that are claimed to be
less than about 10\% (e.g. Schechter et al. 2000). Hence, if the
uncertainty on the value of H$_0$ was {\sl only} due to the
measurement error on the time delay, the technique of gravitational
lensing would already have surpassed that of the local distance-ladder
techniques in accuracy, which in case of the HST Key-Project is about
10\% in their final value of H$_0$=72$\pm$8~km\,s$^{-1}$\,Mpc$^{-1}$
(e.g. Feedman et al. 2001).  Unfortunately, however, not the
measurement of the time delays, but the determination of the
deflector potential\footnote{The `deflector potential' includes all
gravitational effects by which a photon can deviate from its global
geodesic, which assumes homogeneity and isotropy of the universe
(i.e. the FRW universe).} is at present the `bottle-neck' in the attempt 
to accurately determine the value of H$_0$ from GL systems.

To solve the latter problem, it is clear that one would like to have a
significantly larger sample of GL systems with measured time delays
than is currently available. This will (i) reduce the statistical
error on the average value of H$_0$ inferred from different GL
systems, which is dominated by the errors on the measured time delays,
(ii) allow one to select only those GL systems for the determination
of an average value of H$_0$ that have `clean' surrounding fields and
(iii) enable one to find systematic differences between GL systems for
example due to differences in the slope of the radial mass profile or
the mass-sheet degeneracy. Unfortunately, systematic uncertainties in
the deflector potential (e.g. the slope of the radial mass profile)
could potentially `skew' values of H$_0$, determined from different GL
systems, in the same direction. Hence, even though the resulting
statistical scatter can be relatively small (e.g. Koopmans \&
Fassnacht 1999), a large systematic uncertainty (i.e. a scale-factor
in H$_0$) can remain undetected. This problem can only be solved with
detailed modeling of each individual GL system, making use of all
available information such as extended image structure (e.g. rings,
arcs, jets), knowledge about the lens potential (e.g. the stellar
velocity dispersion in the lens galaxy) or general ideas about the
structure of galaxies (e.g. N-body simulations, rotation curves).  
Not all GL systems have this additional information
readily available, however, which again stresses the need to increase
the number of GL systems with measured time delays.

For this reason, the {\sl Cosmic Lens All-Sky Survey} (CLASS)
collaboration (e.g. Browne \& Myers 2000) has started to monitor a 
number of GL systems over the
past few year. In Sect.2, I will review results from three systems
with measured time delays. In Sect.3, I shortly discuss the values of
H$_0$ estimated from these GL systems, under some very simple
assumptions. In Sect.4, I discuss future prospects, including two new
programs with the {\sl Very Large Array} (VLA) and {\sl Multi Element
Radio-Linked Interferometer Network} (MERLIN) to monitor a combined
total of 15 CLASS GL systems.

\section{Time Delays from CLASS Gravitational Lenses}

\paragraph{B0218+357}

The GL system B0218+357 was discovered (e.g. Patnaik et al. 1993) as
part of the {\sl Jodrell Bank-VLA Astrometric Survey} (JVAS), which is
the brighter subsample (i.e. $S_{\rm 5\,GHz}$$\ge$200\,mJy) of the
CLASS survey.  The system consists of two lensed images of a
flat-spectrum radio core, separated by 0.335 arcsec, and an Einstein
ring that results from more extended steep-spectrum source structure.
The redshift of the source is 0.96, whereas the deflector (a
relatively isolated spiral galaxy) has a redshift of 0.68. Corbett et
al.  (1996) reported a time delay of 12$\pm$3\,d (1--$\sigma$
error). More recently, Biggs et al. (1999) presented the results from
a VLA A--array monitoring campaign. From the percentage linear
polarization, polarization angle and 8.5 and 15--GHz flux-density
light curves, a time delay of $\Delta t_{\rm B-A}$=10.5$\pm$0.4\,d
(95\% confidence) was measured. This value was confirmed by Cohen et
al. (2001), who find $\Delta t_{\rm B-A}$=10.1$^{+1.5}_{-1.6}$\,d
(95\% confidence), using independent data obtained with the VLA during the
same period as Biggs et al.

\paragraph{B1600+434} 

The GL system B1600+434 (Jackson et al. 1995) consists of two compact
flat-spectrum radio images, separated by 1.39 arcsec, of a quasar at a
redshift of 1.59. The primary lens galaxy is an edge-on spiral galaxy
at a redshift of 0.41 (Jaunsen \& Hjorth 1997; Koopmans et al. 1998).
An A and B--array VLA 8.5-GHz monitoring campaign gave a time delay of
$\Delta t_{\rm B-A}$=47$^{+12}_{-9}$\,d (95\% confidence) (Koopmans et
al. 2000). More recently, a value of $\Delta t_{\rm B-A}$=51$\pm$4\,d
(95\% confidence) was found from an optical monitoring campaign with
the {\sl Nordic Optical Telescope} (NOT) (Burud et
al. 2000). Preliminary results from a new multi-frequency monitoring
campaign with the VLA seem to confirm these results.

\paragraph{B1608+656}

The GL system B1608+656 consists of four compact flat-spectrum radio
images with a maximum image separation of 2.1 arcsec (Myers et
al. 1995). The source has a redshift of 1.39 and is being lensed by
two galaxies inside the Einstein radius, of which at least the
brightest has a redshift of 0.63. In the optical and near-infrared the
host galaxy of the radio source is lensed into prominent arcs (Jackson
et al. 1998). Fassnacht et al. (1999) have measured all three
time delays from radio light curves obtained in 1996--1997 at 8.5 GHz 
with the VLA in A and B--array. Combined with data from a similar
campaign in 1998, their preliminary results are: $\Delta t_{\rm
B-A}$=26\,d, $\Delta t_{\rm B-C}$=34\,d and $\Delta t_{\rm
B-D}$=73\,d, with an error of 5\,d (95\% confidence) on each time delay
(Fassnacht et al. 2000).

\section{Estimates of the Hubble Constant}

To estimate the value of H$_0$ from these time delays requires a good
model of the deflector potentials. In all three GL systems, it is
{\sl assumed} that these are dominated by the potential of the primary
lens galaxies (two in the case of B1608+656) and that these galaxies 
have an {\sl isothermal} mass distribution.

Under these assumptions (see the references for more details) one
finds: H$_0$=$69^{+13}_{-19}$~km\,s$^{-1}$\,Mpc$^{-1}$ (95\%) from
B0218+357 (Biggs et al. 1999),
H$_0$=$60^{+15}_{-12}$~km\,s$^{-1}$\,Mpc$^{-1}$ (95\%) from B1600+434
(Koopmans et al. 2000) and
H$_0$=$63^{+7}_{-6}$~km\,s$^{-1}$\,Mpc$^{-1}$ (95\%) from B1608+656
(Koopmans \& Fassnacht 1999) with $\Omega_{\rm m}$=0.3 and
$\Omega_{\Lambda}$=0.7. Burud et al. (2000) estimate a slightly lower
value of H$_0$ from B1600+434, using the mass models from Maller et
al. (2000). In addition, Lehar et al. (2000) claim a larger systematic
error on the value of H$_0$ from B0218+357, due to the uncertainty in
the position of the lens galaxy. Result from modeling the Einstein
ring in B0218+357 (Wucknitz, private communications) seem to agree
with the galaxy position used by Biggs et al, however. Although we
stress that these are preliminary values, the interesting conclusion
from a comparison of these values of H$_0$, is their good agreement
not only with determinations from other GL systems, but also with
those from the HST Key-Project, S--Z measurements and determinations
from high-redshift SNe Ia (e.g.  Koopmans \& Fassnacht 1999).

Even so, the uncertainties are still considerable and not all possible
mass models have been fully explored yet.  In none of these cases for
example does the error include the uncertainty in the slope of the
radial mass profile or the center of the mass distribution, which
dominate the systematic uncertainties in the value of H$_0$.  To
improve this situation, Wucknitz et al. are using the additional
information in the structure of the radio Einstein ring in B0218+357
to constrain the position of the lens galaxy and its radial mass
profile. Similarly, Surpi \& Blandford are using the arcs in B1608+656
to further constrain its mass distribution, whereas Fassnacht et
al. have obtained data to measure the central velocity dispersion of
the primary lens galaxy. In the case of B1600+434, no clear extended
source structure is present, although Keck observations will be done
to try to measure the velocity dispersion and rotation velocity of the
bulge and disk, respectively.

\section{The Future of H$_0$ from Gravitational Lensing}

For the three CLASS GL systems discussed above (Sect.2), the time
delays are or will soon be known with errors much less than 10\%. With
the ongoing effort to improve the determination of the lens potentials
of, in particular, B0218+357 and B1608+656 (Sect.3), one can also
expect the uncertainty on the inferred time delays to reduce to less
than 10\% in the near future.  These three systems will then give an
average global value of H$_0$ comparable in accuracy to the results
from the HST Key-Project. Together with other GL systems that have
measured time delays, this situation can only improve. Another example
of a very promising CLASS GL system is B1933+503, for which the
inferred time delay from mass modeling has an uncertainty $\le$15\%,
with excellent opportunities for improvement (Cohn et al. 2000; see
also Nair 1998).  Although no time delay could be determined from a
VLA monitoring campaign (Biggs et al. 2000), the source has in the
past varied by as much as 33\% at 15 GHz and is currently being
re-observed with both the VLA and MERLIN.  To increase the number of
GL systems with measured time delays, CLASS is now engaged in two new
monitoring projects with the VLA (8 systems; PI: Fassnacht) and MERLIN
(Key-Programme; 12 systems; PI: Koopmans). In total 15 CLASS GL
systems will be monitored (including those in Sect.2). With ongoing
optical monitoring programs, the total number of GL systems being
monitored in 2001 will likely be 20--30\,!  Although not every systems
will yield time delays, we expect that the number of GL systems with
measured time delays is likely to double in the next few years.

However, in order to obtain a `competitive' global measurement of
H$_0$ from gravitational lensing, {\sl the focus in the coming years
needs to be on improving the determination of the deflector potential
of each individual GL system} from which time delays are being
measured. From the work being done at present, this appears an
attainable goal. In light of the fact that the first GL system was
discovered over twenty years ago, progress might appear slow. However,
the first unambiguous measurement of a time delay was done only some
five years ago and since then at least seven GL system have been added
to this list, some of them having much simpler deflector potentials
than Q0957+561, which has received the most attention over the last
two decades.

Finally, we can ask ourselves the question: Is it still worthwhile to
measure H$_0$ from gravitational lensing, now that the HST Key-Project
has determined the local value with an uncertainty of around 10\%\,?
Here, one should keep in mind that the value of H$_0$ determined from
gravitational lensing is a `global' determination, whereas that
determined from the HST Key-Project is a `local' value. Both methods
are therefore in some sense complementary and do not necessarily have
to result in the same value for the expansion speed of the universe
(i.e. locally H$_0$ could differ from its global average value), even
though this is often implicitly assumed based on the idea that the
universe is homogeneous and isotropic on large scales (but not
necessarily on small scales!). The latter results in a global (R--W)
metric and a set of global parameters describing its evolution, which
by definition has the same local and global value of H$_0$.
Agreement or disagreement between values of H$_0$ from two or more
independent and different methods can therefore elucidate our
understanding of the universe and in case of agreement put its
determination on a much firmer basis.


\section*{Acknowledgements}

LVEK thanks the organizers of the OzLens workshop for a stimulating
meeting and partial support.

\section*{References}

\reference Biggs, A.\ D., Browne, I.\ W.\ A., Helbig, P., Koopmans,
L.\ V.\ E., Wilkinson, P.\ N.\ \& Perley, R.\ A.\ 1999, MNRAS 304,
349

\reference Biggs, A.\ D., Xanthopoulos, E., Browne, I.\ W.\ A.,
Koopmans, L.\ V.\ E.\ \& Fassnacht, C.\ D.\ 2000, MNRAS 318, 73

\reference Browne, I.\ W.\ A.\ \& Myers, S.\ T.\ 2000, IAU Symposium, 201, E47 

\reference Burud, I.\ et al.\ 2000, ApJ 544, 117 

\reference Cohen,  A.\ S.\, Hewitt, J.\ N.\ , Moore, C.\ B.\, Haarsma, D.\ B.\
2001, ApJ in press, astro-ph/0010049 

\reference Cohn, J.D., Kochanek, C.S., McLeod, B. A., Keeton, C.R. 
2000, submitted, astro-ph/0008390

\reference Corbett, E.\ A., Browne, I.\ W.\ A.\ \& Wilkinson, P.\ N.\
1996, IAU Symp.\ 173: Astrophysical Applications of Gravitational
Lensing, 173, 37

\reference Fassnacht, C.\ D., Pearson, T.\ J., Readhead, A.\ C.\ S.,
Browne, I.\ W.\ A., Koopmans, L.\ V.\ E., Myers, S.\ T.\ \& Wilkinson,
P.\ N.\ 1999, ApJ 527, 498

\reference Fassnacht, C. D. , Xanthopoulos, E., Koopmans, L. V. E.,
 Pearson, T. J., Readhead, A. C. S., Myers, S. T., 1999, astro-ph/9909410,
 to appear in
 "Gravitational Lensing: Recent Progress and Future Goals",
 T. G. Brainerd and C. S. Kochanek (eds.)

\reference Freedman, W.\ L.\ et al.\ 2001, ApJ in press,
astro-ph/0012376

\reference Jackson, N.\ et al.\ 1995, MNRAS 274, L25 

\reference Jackson, N., Helbig, P., Browne, I., Fassnacht, C.\ D., 
Koopmans, L., Marlow, D.\ \& Wilkinson, P.\ N.\ 1998, A\&A 334, L33 

\reference Jaunsen, A.\ O.\ \& Hjorth, J.\ 1997, A\&A 317, L39 

\reference Koopmans, L.\ V.\ E., de Bruyn, A.\ G.\ \& Jackson, N.\ 1998, 
MNRAS 295, 534 

\reference Koopmans, L.\ V.\ E.\ \& Fassnacht, C.\ D.\ 1999, ApJ 527,
513

\reference Koopmans, L.\ V.\ E., de Bruyn, A.\ 
G., Xanthopoulos, E.\ \& Fassnacht, C.\ D.\ 2000, A\&A 356, 391 

\reference Kundic, T.\ et al.\ 1997, ApJ 482, 75

\reference Leh{\'a}r, J.\ et al.\ 2000, ApJ 536, 584

\reference Maller, A.\ H., Simard, 
L., Guhathakurta, P., Hjorth, J., Jaunsen, A.\ O., Flores, R.\ A.\ \& 
Primack, J.\ R.\ 2000, ApJ 533, 194

\reference Myers, S.\ T.\ et al.\ 1995, ApJL 447, L5

\reference Nair, S.\ 1998, MNRAS 301, 315 

\reference Patnaik, A.\ R., Browne, I.\ W.\ A., King, L.\ J., Muxlow,
T.\ W.\ B., Walsh, D.\ \& Wilkinson, P.\ N.\ 1993, MNRAS 261, 435

\reference Refsdal, S.\ 1964, MNRAS 128, 307 

\reference Schechter, P.\ 2000, astro-ph/0009048, to be published in
IAU Symposium No. 201: New Cosmological and the Values of the
Fundamental Parameters, eds. A.N.  Lasenby and A. Wilkinson

\reference Walsh, D., Carswell, R.\ F.\ \& Weymann, R.\ J.\ 1979, Nature 279, 381 

\end{document}